\documentclass[journal]{IEEEtran}

\usepackage{url}
\usepackage{graphicx}
\usepackage{amssymb}
\usepackage{color}

\begin{document}

\title{Towards Decision Support for Smart Energy Systems based on
  Spatio-temporal Models}
\author{
\IEEEauthorblockN{Jan Olaf Blech\IEEEauthorrefmark{1},
Lasith Fernando\IEEEauthorrefmark{1},
Keith Foster \IEEEauthorrefmark{1},
Abhilash G\IEEEauthorrefmark{7}, 
  Sudarsan SD\IEEEauthorrefmark{7}} \\
\IEEEauthorblockA{\IEEEauthorrefmark{1} RMIT University, Melbourne,
  Australia} \\
\IEEEauthorblockA{\IEEEauthorrefmark{7} ABB Corporate Research, Bangalore, India}
}

\maketitle
\begin{abstract}
This report presents our SmartSpace event handling framework for managing
smart-grids and renewable energy installations. SmartSpace provides
decision support for human stakeholders. Based on different datasources that feed into our framework,
a variety of analysis and decision steps are supported. These decision
steps are ultimately
used to provide adequate information to human stakeholdes. 
The paper discusses potential data sources for decisions around smart energy
systems and introduces a spatio-temporal modeling technique for the involved data. Operations
to reason about the formalized data are provided. Our spatio-temporal models help to provide a semantic context
for the data. 
Customized rules
allow the specification of conditions under which information is
provided to stakeholders. 
We exemplify our ideas and present our
demonstrators including visualization capabilities.
\end{abstract}

\begin{IEEEkeywords} 
collaboration solutions,
decision support,
distributed engineering,
smart-grids,
visualization
\end{IEEEkeywords}

\IEEEpeerreviewmaketitle

\section{Introduction}

Smart energy systems and new solutions for grid technology have become
an important topic in the past decade. The emergence
of renewable energy technology has been a source for new challenges in
academia and industry. 
Research and industrial solutions now cover areas such
as energy storage, strengthening of existing grid technology and
prediction models for energy generation and consumption. The
implementation of these new technologies can require large capital
investments. To support
decisions around the introduction of new smart energy systems and during operation
of existing systems a variety of
databases have been created. In this paper, we are
extending these existing views by integrating the existing data and
knowledge sources that have already been created for smart energy
systems into a common framework. The framework aims at supporting a
variety of human stakeholders in their decisions around the planning,
maintenance and
operation of smart energy systems.

Here, we are discussing  SmartSpace. SmartSpace builds upon and extends our collaborative engineering framework 
which aims at providing information to human stakeholders such as engineers,
operators, managers, and facility owners to support remote operations.
The first versions of
collaborative engineering are primarily focusing on operation and maintenance of remote industrial
facilities such as mines and oil rigs in the Australian outback. 
We have extended our previous work~\cite{smartspace} on
SmartSpace by providing a broader overview on related approaches and
the smart energy context as well as the modeling and reasoning background. We have further extended our semantical modeling and
reasoning that are a core feature of our framework and were able to
incorporate  more feedback on our demonstrators.
Compared to the collaborative engineering framework that has been
published in \cite{etfa,etfa2} we incorporated the following technical
adaptations:
\begin{enumerate}
\item We have ported the  framework from an
  industrial automation and remote Australian mining context
  to the smart energy area.
\item We have tried out new semantical modeling elements, developed new operators and integrated them into our reasoning framework.
\item We provide a discussion on the domain specific data sources for the smart
  energy sector.
\item We have realized a demonstrator: SmartSpace with SmartSpace3D, a
  visualization front-end.
\end{enumerate}
Our demonstrator
comprises the following ingredients:
\begin{itemize}
\item It connects to external data-sources.
\item It features, a spatio-temporal reasoning component.
\item It comprises, a visualization layer and user interface.
\end{itemize}
Our SmartSpace framework provides decision support for
human stakeholders such as customers / consumers, engineers, owners,
and operators of
smart energy systems. We have identified two different ways to support
decisions: 
\begin{itemize}
\item
We support {\it short-term decisions}, such as preventing a grid blackout due to
failure, maintenance, environmental conditions (photo-voltaic and
wind).
\item
{\it Longer-term decisions} such as strengthening an
electricity grid permanently are also supported.
\end{itemize}
Longer-term and short-term decisions both aim at supporting humans in their decisions.

\subsection{Motivation: A Smarter Distribution Network}

\begin{figure*}
\centering
\includegraphics[width=.75\textwidth]{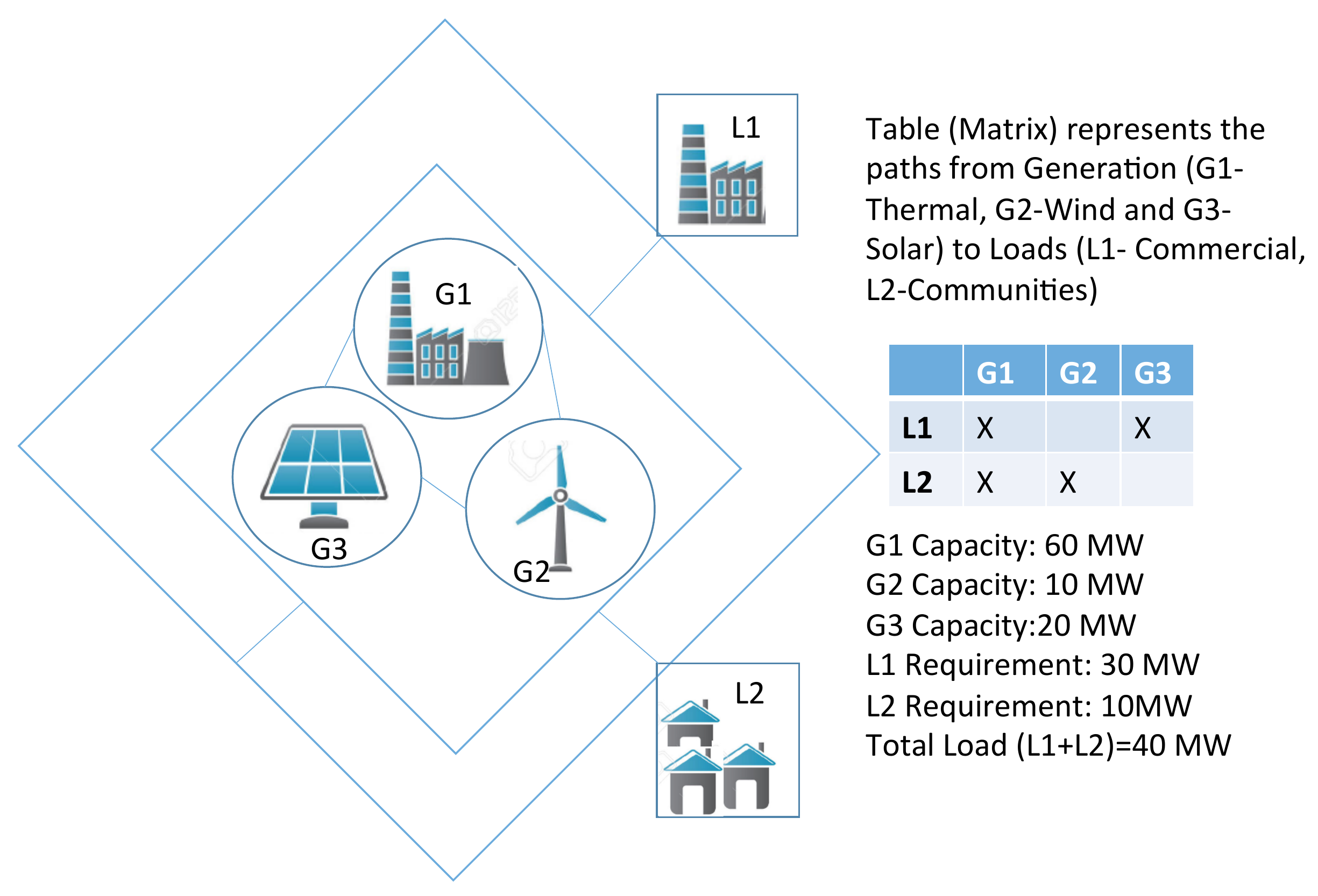}
\caption{Smart Grid Example}
\label{fig:sg}
\end{figure*}

Figure~\ref{fig:sg} gives an overview on a new scenario of distributed
electricity loads and paths to get there: a meshed network. Here, a coal fired plant, solar plant and wind power plant are connected. The connections establish network paths
in the distribution network. The management of loads and generation
capacity is encoded as a matrix. This is a simple form of embedding such intelligence, so that one can identify the paths available and accordingly
improve the availability of the system by also trading from the
network in a
coordinated manner. This information can be formalized using a
spatio-temporal model to make it accessible for machine-based processing.

We are interested in remotely monitoring, co-ordinating and operating 
components involved in electricity generation and grid technology. A
distribution network capable of reduced losses, improved voltage
regulation and control management of
reactive power is the key to this improvement.

With an increasing number of distributed energy resources  including renewable sources, distributed architectures need
improvement. Two key performance indicators, System Average Interruption Duration Index (SAIDI) and Customer
Average Interruption Duration Index (CAIDI) are increasingly of significance from the system point of view. The
challenge is dealing  with this increased complexity in a distributed system where the fine balance of coordinated or collaborative decision
making across centralized elements like SCADA and decentralized aspects of controllers and protection relays is a
necessity. Many protection and control elements in distribution automation use constrained processing
power ranging from 8 bit microcontrollers to more advanced 32 bit
ones. 
The need for intelligent decisions over smart electricity
network is increasingly evident in the need for automatic operations
in normal load conditions, as well as during Fault restoration
Fault Detection, Isolation and Restoration (FDIR). FDIR is a major topic of discussion in emerging scenarios. There are
already provisions in electricity distribution automation to handle FDIR. Still the smartness of these systems need significant
uplift to handle the diverse challenges. Also with new technologies of data analysis and inference, these challenges
can easily be translated into opportunities \cite{sgbook}.
Some decision support in smart energy systems comes with real-time
constraints: In the past  manual intervention was required to open fuse and close signals to energize circuit breakers, today
it is possible to have programmed reclosing. We require a short time
between the time of detection of fault to its isolation to avoid a
system failure. At a high level, appropriate models for a decentralized FDIR attempt are a
connecting matrix to represent the distribution network  scheme which includes three types of components: sources,
switching devices, and loads. The FDIR logic reads load current of each device, and updates real time loading for each
load.  The logic searches all possible paths from source to open tie re-closer and calculates the expected load on
new paths. It sends a load message to a protection relay which in turn changes its settings. This would  lead to 
sending a close command to Tie -Point recloser. Once the fault is
cleared, the FDIR logic is able to restore the network
to pre-fault condition by reversing FDIR actions and safely checking
all conditions. 
The FDIR logic  can have elements organized centrally in a cloud or
in a SCADA system and also elements distributed across the system elements including relays. This collaborative control
 necessitates the need of collaborative smart  decision making and
 also consideres decentralized real-time aspects of the
system.

\subsection{Overview on SmartSpace and Collaborative Engineering} 

\begin{figure}
\centering
\includegraphics[width=.4\textwidth]{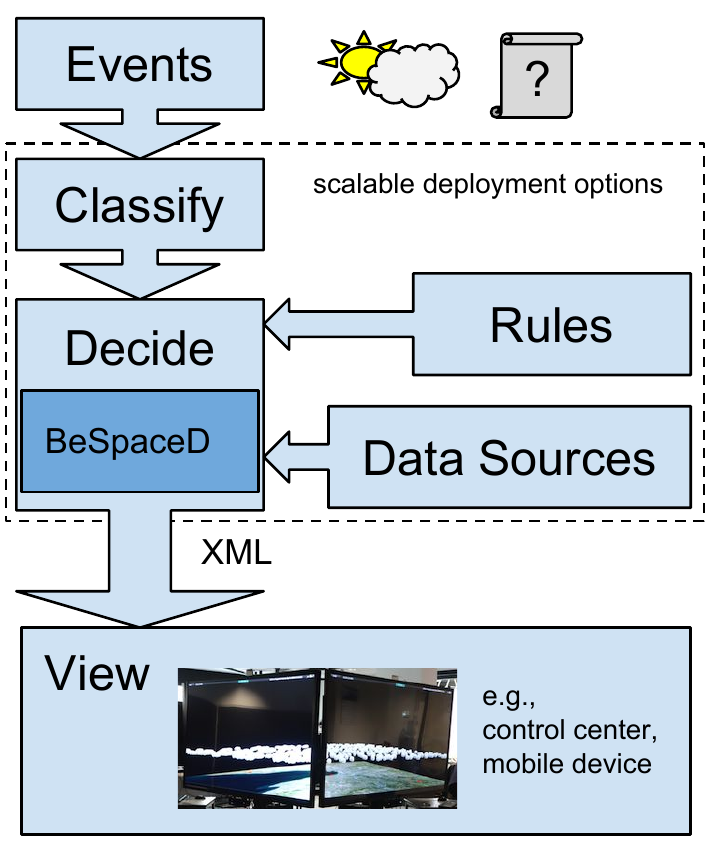}
\caption{Overview of the SmartSpace framework}
\label{fig:overview}
\end{figure}

Figure~\ref{fig:overview} presents an overview of the steps involved
in our existing {\it collaborative engineering} decision support
framework. We follow the description from previous
work (\cite{etfa,etfa2}).
Decision support is triggered by events which can be diverse in
nature:
\begin{itemize}
\item
 They may be manually triggered, originating from
humans. Examples comprise a consulting request to assess the quality of a grid
infrastructure or an on-site engineer reporting to a service. In this
case only a few events per week, month or year may
arrive. 
\item
On the other hand, they can come from machines. For example, events can also be associated with alarms
originating from PLCs. Thousands of events per second may
arrive \footnote{Such events can be used for multiple purposes. In addition to our collaborative engineering framework, we are using them for monitoring of software behavior \cite{wenger1,wenger2}.}. 
\end{itemize}
Collaborative engineering responds to the events by using customizable
decision support that is based on our spatio-temporal reasoning
framework BeSpaceD \cite{bespaced0,bespaced1}. XML code is generated and send to different
devices for different stakeholders. The devices themself interpret the
generated code, thereby triggering the display of relevant, typically
stakeholder, location and device specific information.

Our collaborative
engineering framework is scalable, it may be deployed in a massive parallel cloud environment or
on a single laptop computer. The decision support comprising the BeSpaceD
instantiations may also run on multiple virtual machines in parallel
when serving different events.

In the smart energy systems case, we have a combination of both (i) longer term
human triggered events, such as consulting requests and (ii) short term
machine generated events, such as
alarms in an automation facility or
updates in weather data (for some forms of renewable energy). The decision support component is
instantiated by customizable rules that encode the business
logic. Rules can rely on additional data sources. Based on this, a
response, view or recommendation is created. In case of smart energy
systems, the XML encoded result is
interpreted for multiple form-factors, e.g., a mobile device of an
on-site technician or a command center.  Such information can comprise
the display of websites, control screens, camera views, annotated pictures, start of
software or
interactive models as shown at the bottom of the figure.

Collaborative engineering and its SmartSpace decision support
instantiation is realized using rules, encoded as customized
components. In discussions with stakeholders, we have identified the
following goals with respect to operation:
\begin{enumerate}
\item
{\it Classification and prioritization of events.} In the case, where we are dealing with a large number
of events arriving in a short time interval, human operators may be
overwhelmed by coping with the details of many events. We are interested in
providing ways for domain experts to customize the priority and
presentation of event triggered information.
\item {\it Customization of decision support rules.}  Human
  stakeholders may need
  different information based on their roles and locations. For example, a service technician on-site may
  be concerned with the immediate replacement of a damaged solar
  panel. On the other hand support staff in a service center may be interested in gathering data on
  failure modes and rates. Our decision support rules allow the
  stakeholder specific filtering and processing of information as well
  as the generation of stakeholder specific responses such as visualizations.
\item {\it Runtime Adaptation}. Adding and removing of decision support rules on the fly in the running
  system is a goal of our framework. For example, decision support systems must be able to adapt
  to changes in the infrastructure, such as being able to cope with
  new power lines and power plants as well as new data sources. This
  may significantly change the required recommendations by the
  system, but changes should only have a small impact on the IT
  infrastructure of the running system.
\end{enumerate}

\subsection{Example: Short-Term Decision Support for PV Panels}
An  example for short-term decision support comprises a
platform for  providing information including visualization of
relevant information based on live-weather data. The live weather is
streamed to our SmartSpace framework and analyzed. Rules are used to
select relevant stakeholders and
trigger the display of relevant notifications and processed
information to them.
The data analysis can reveal and predict shortages of electricity production in PV panels due
to excessive rain-cloud coverage. 

\subsection{Example: Identification of Weak-links}
\begin{figure}
\centering
\includegraphics[width=.475\textwidth]{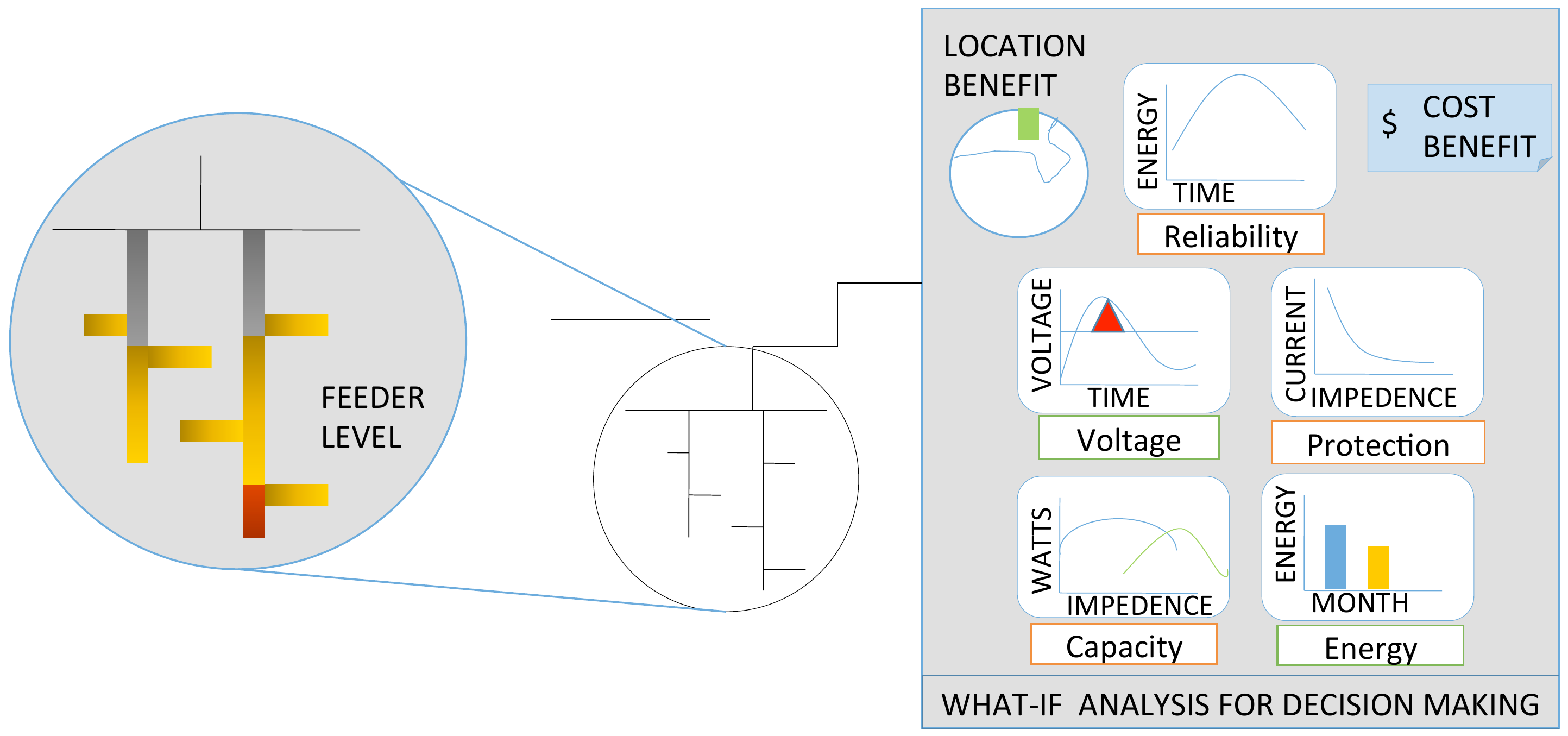}
\caption{Identifying weak links in an electricity grid}
\label{fig:wl}
\end{figure}

The second motivating example explains longer-term decision support. It comprises the
identification of weak-links in electrical grids.
The assessment of electrical grids  in terms of the strength of the
network and the identification of the weak links \cite{pes2015} is a major
topic of interest around the world. One can take a look at the current balance between
existing electricity generation,
future requirements for electricity generation and electrical
loads. This is location dependent and may depend on the time of day.
Various data sources can be combined for this analysis. This can  lead
to spatio-temporal models and a heat 
map like the one shown in Figure~\ref{fig:wl}.  
A plan to overcome the identified weak links can be
developed once the analysis is concluded. One may consider the possibility of having 
various renewable energy sources distributed power generation or microgrids to strengthen the weak 
links. If we are analysing an electricity grid on an island,  there is also possibility
that the local microgrid is totally 
isolated from a main grid.  
The analysis for strengthening the weak links involves  the
consideration of 
electrical system across multiple parameters of energy needs, as well as
reliability, voltage, and current as shown in Figure~\ref{fig:wl}.

\subsection{Overview}
The paper is structured as follows: We discuss related work in
Section~\ref{sec:relwork} and introduce our formal modeling and reasoning framework
 in Section~\ref{sec:bespaced}. The modeling and reasoning section also
features the presentation of  new operators and modeling constructs
for our SmartSpace framework. Section~\ref{sec:data}
discusses relevant data sources for decision support in the smart grid
area and Section~\ref{sec:demo} presents our demonstrator. A
conclusion is featured in Section~\ref{sec:concl}.

\section{Related work}
\label{sec:relwork}
An overview on challenges and needs for Smart Grid systems is provided
\cite{sg}. A variety of software approaches and related requirements
on the networking and IT infrastructure \cite{gungor} to support different aspects of
smart grids and smart energy management have been established. 
Different techniques for demand side management have been outlined in \cite{dsm}.
Small grid sizes such as the electricity grid within an office building
has been considered in \cite{fortiss}. 
Security and robustness can be an issue in smart energy systems, this
has been discussed in \cite{amin}. Furthermore,  operational
needs for different energy sources need consideration. A focus on photo-voltaic operations
relevant for our paper is studied in \cite{pv}. 

We complement the research done on  smart grids/smart energy
systems by concentrating on software support to manage these
systems. in our work, we provide automatic decision support for human operators
in a control room, or in the field by using a mobile form-factor. We
build on the
collaborative engineering framework as explained in the introduction and use a
decision support component based on spatio-temporal models and
reasoning. For this reason, our work is  related to existing collaboration
software in the industrial space such as Dassault Syst{\`e}me's
Enovia~\cite{enovia} and Delmia~\cite{delmia} that focus on the
visualization of production plants and allow the association of data
with elements in the visualization.
In addition, our framework relies on reasoning about semantic models
and for this reason
ontology-based approaches are also related. For example, existing approach can be based on semantic
web technology (see
\cite{sure}). Related to this, collaboration issues are  studied in the
ComVantage  \cite{salmen} project with a focus on a
mobile enterprise reference framework for future internet
operability. Some applications in the industrial automation
context are featured. Another framework for collaborative requirements engineering is
presented in \cite{c-farm}. 

Different approaches have been introduced for spatio-temporal modeling
and reasoning. For describing models using logic work on process
algebra-like formalisms has been conducted \cite{cardelli04,haar}.  Qualitative spatial
descriptions and reasoning \cite{Bennett,randell,cohn,chen} where spatial relationships are not
described through exact geometry but rather using predicates for,
e.g., describing that an entity is located close to another object are
important for the BeSpaceD based abstractions. Other logic
approaches to spatial reasoning can be found in
\cite{hirschkoff,zilio}.
A comparison on semantic formalisms
for industry 4.0, as well as guidelines to assist engineers on this
topic is featured  in~\cite{chihhong}. Related to our BeSpaceD
operators introduced in this paper, functional programming language features for
large scale data operations are
supported by frameworks such as
Spark\footnote{\url{http://spark.apache.org/}}. In comparison, in our work, we are
specifically targeting operations for spatio-temporal models.

\section{BeSpaceD}
\label{sec:bespaced}
Means for automatically analysing information and computing results
for display to human stakeholders is a core functionality of our
framework. We base our decision support for smart energy systems on
spatio-temporal models that represent ontologies and data thereby
assigning a semantical meaning to data and establishing relations
between different artefacts.

BeSpaceD is our spatio-temporal modeling and reasoning framework: it
comes 
as a general purpose modeling and reasoning framework. This means that
applications are not limited to the smart-energy context.
In the work presented in this paper, we use our BeSpaceD framework as:
\begin{itemize}
\item a description language, for achieving formal models of
  facilities such as power plants,
  grids and relations between entities;
\item a format to store spatio-temporal data;
\item a way to reason about the formalized models and data by using BeSpaceD's libraries.
\end{itemize}
In the following,
we describe the modeling and data formalization language and the BeSpaceD-based reasoning
library functionality. In addition, we 
provide some implementation background.

The BeSpaceD framework is implemented in the Scala programming language which is
bytecode compatible with Java. This means, BeSpaceD's core functionality runs in a
Java environment. On one hand BeSpaceD-based services can be offered as
cloud services using highly scalable infrastructure, on the other hand
one can
also run BeSpaceD locally on embedded devices that support a Java runtime
environment. In addition to decision support systems for industrial
automation and smart energy systems, we have successfully applied BeSpaceD
for coverage analysis in the area of mobile devices~\cite{han2015} and for verification of spatio-temporal
properties of industrial robots~\cite{apscc,fesca2014,cyberbeht}. BeSpaceD has been integrated into a model-based development approach \cite{hordvik} and models of industrial operations have been part of our work \cite{harland} as well as applications to adaptive mobile systems \cite{taherkordi}.

\subsection{Modeling in BeSpaceD}

Models in BeSpaceD are formalized as abstract data types. These are
realized in BeSpaceD as Scala case classes which can be combined to
form larger constructs to represent a model. The case classes are
compatible with Java and ``live'' in the java runtime
environment. Overall, the case class-based formalization provides a
functional programming language feeling.
To provide a look and feel, some of our language constructs (see also \cite{2015arXiv151204656O})
are provided in Figure~\ref{fig:belang}.
At the core of our modeling language, an {\tt Invariant} is the basic
logical entity and is
supposed to hold for a system. Despite this, invariants
can and typically do, contain conditional parts. For example, a logical formula
can be part of the overall invariant, but  require a precondition to
hold, such as a point in time implying a
 state of a system. 
Following the abstract data type concept, constructors for basic logical operations connect invariants to form a
new invariant. 

\begin{figure*}
{\small

\textsf{Examples for basic logic operators}
\begin{verbatim}
case class OR (t1 : Invariant, t2 : Invariant) extends Invariant;
case class AND (t1 : Invariant, t2 : Invariant)  extends Invariant;
case class NOT (t : Invariant) extends Invariant;
case class IMPLIES (t1 : Invariant, t2 : Invariant)  extends Invariant;
case class TRUE() extends ATOM;
case class FALSE() extends ATOM;
\end{verbatim}
\textsf{Examples for time predicates}
\begin{verbatim}
case class TimePoint [T] (timepoint : T) extends ATOM; 
case class TimeInterval [T](timepoint1 : T, timepoint2 : T) extends ATOM; 
\end{verbatim}
\textsf{Events and Ownership}
\begin{verbatim}
case class Event[E] (event : E) extends ATOM;
case class Owner[O] (owner : O) extends ATOM;
\end{verbatim}
\textsf{Examples for spatial predicates}
\begin{verbatim}
case class OccupyBox (x1 : Int,y1 : Int,x2 : Int,y2 : Int) extends ATOM;
case class Occupy3DBox (x1 : Int, y1: Int, z1 : Int, 
             x2 : Int, y2 : Int, z2 : Int) extends ATOM;
case class OccupyPoint (x:Int, y:Int) extends ATOM
\end{verbatim}
\textsf{Topology, Graphs}
\begin{verbatim}
case class Edge[N] (source : N, target : N) extends ATOM 
case class Transition[N,E] (source : N, event : E, target : N) extends
ATOM 
\end{verbatim}
}
\caption{Excerpt of logical operators for the BeSpaceD language}
\label{fig:belang}
\end{figure*}

In the Figure~\ref{fig:belang}, 
the first part provides operators from propositional logic  (e.g., {\tt
 AND}, {\tt OR}). The second
part provides predicates for time such as {\tt TimePoint} and {\tt
  TimeInterval}. 
The third part allows the
specification of events and ownership of logical entities. Ownerships
are used to associate time and space with specific entities. The
fourth part provides constructs for space, in particular geometry. 
The {\tt OccupyBox} predicate refers to a rectangular two-dimensional geometric
space which is parameterized by its left lower and its right upper
corner using a cartesian coordinate system.  The fifth part
provides constructs for the specification of mathematical graphs using
edges and
state transition systems using transitions. In the design of domain specific languages, the
question whether  to make a construct a {\it first class citizen} of the
language, e.g., by providing a separate constructor or to integrate it
into another construct (e.g., as a parameter) is a challenging
decision. In our work, we are concentrating on spatio-temporal aspects and
the concept of ownership in space and time. We found this useful in
the industrial automation and smart energy domains. The choice of our language
constructs simplify the translation and validation of requirements
dealing with time and space into BeSpaceD.

A concrete BeSpaceD example is provided below. The following BeSpaceD formula expresses that the
rectangular two-dimensional geometric space with the upper-left corner
point $(145,4056)$ and the lower-right corner point $(1536,
2609)$ is subject to a semantic condition {\tt AreaOfImpact} between
integer-defined time points $t1$ and $t2$.

{\small
\begin{verbatim}
...
  IMPLIES(AND(TimeInterval(t1,t2),
               Owner("AreaOfImpact")),
      OccupyBox(145,4056,1536,2609))
...
\end{verbatim}
}

The {\tt AreaOfImpact} is a semantic condition using the ownership predicate and can be replaced
with a more concrete identifier, for
example by an indicator for a rain-cloud or a solar panal on a
weather map. 

By combining the different constructors, BeSpaceD formula can be
constructed to formalize relevant information for our
demonstrators. Examples include
average UV intensity in an area, specification of capacity and
location of power-lines, typical power consumption and generation in
an area (parameterized by time). Specifications can get long and can
easily comprise a few 1000 interconnected data-elements. Data can be imported while the system
is running. This enables the possibility to process live streamed data
and integrate it into the decision process. Although the Java-based
memory representation of case classes can be efficient, for import and
export and optimization of
processing time, we have developed different JSON and XML-based
formats for storing the data.

\subsection{Reasoning in BeSpaceD}
While models describe systems and data,  we also need ways to retrieve
information from them, deduct implications, thus reason
about them.
For this reason, BeSpaceD comprises library-like functionality to handle specifications. 
Our  library provides means for the efficient analysis of BeSpaceD
formulas: Among other aspects, the
functionality comprises, ways of abstraction, filtering and efficient
processing of information. Since BeSpaceD's focus is on spatio-temporal
reasoning, a variety of operators filter and abstract time and
space. An example of breaking down a complex task to several simpler
tasks is the transformation of geometric constraints on
areas into  geometric constraints on points. Several implementations
exist in BeSpaceD, each one has different operational characteristics. 

\subsection{Interacting with the Runtime System}
BeSpaceD needs to interact with its runtime system for importing and
exporting data and interacting with other tools and parts of the SmartSpace Framework.
Various ways to import and
visualize information are supported. Examples comprise the import and
export of information
from/into databases, but also the collection of information from sensors
and the conversion into BeSpaceD data structures. This can rely on
streaming. 

In addition, we have connected tools such as an external
SMT solvers (e.g., we have a connection to z3 \cite{z3}) for external
processing of some information. SMT solvers can help to resolve geometric
constraints such as deciding whether there is an overlapping of different areas in time and space.
Functionality that deals with filtering of datatypes, normalization and spatio-temporal
fold operations are especially critical for our SmartSpace
framework. Some of this is available in general BeSpaceD framework,
but the folding functionality has been implemented specifically
with SmartSpace in mind.

The connection to the SmartSpace and collaborative engineering
framework is typically done by automatically creating an individual Java virtual
machine with a running BeSpaceD framework for each decision support
request and retrieving the XML based results.

\subsection{Aggregating Information in Space and TIme}
\begin{figure}
\centering
\includegraphics[width=.475\textwidth]{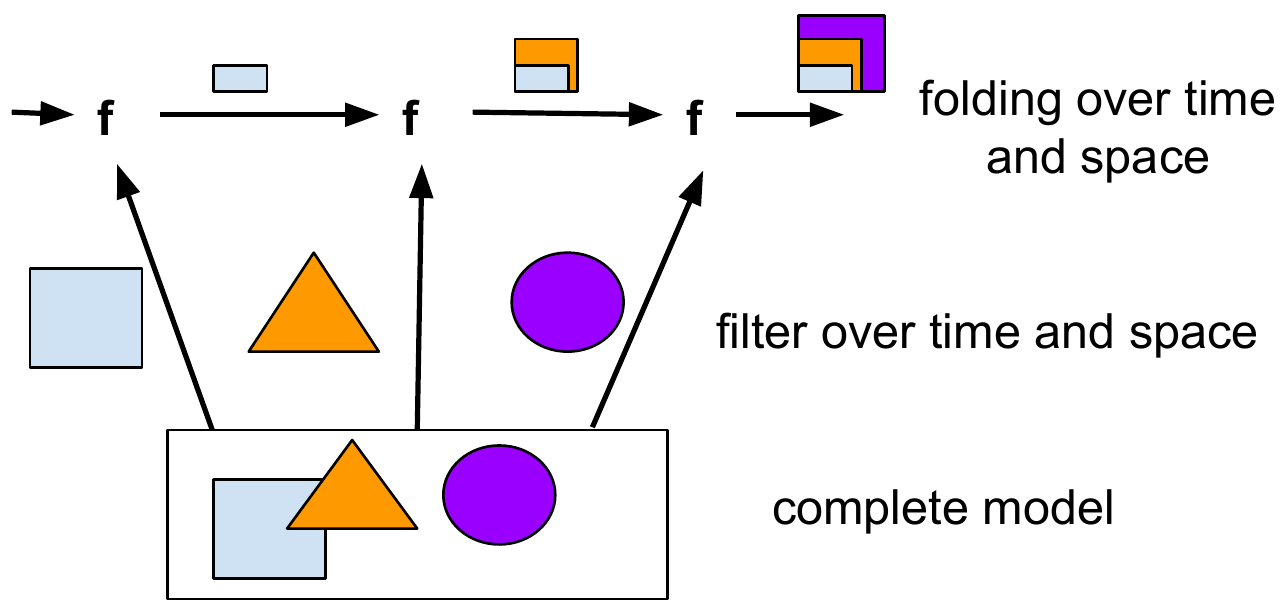}
\caption{Processing spatio-temporal information using an aggregating fold
  operation \cite{newoperators}}
\label{fig:spacetimefold}
\end{figure}

Operations aggregating information over space and time have been found
useful for our smart energy systems work. The operators are described
in detail in a technical report \cite{newoperators}.
Figure~\ref{fig:spacetimefold} provides an abstract view of the aggregation of spatio-temporal data using fold operations.
The so called {\it folding} aggregates data from larger data-structures by iterating over
time and space intervals. Each iteration step looks at one particular
 interval. This is represented in the figure as
a circle, square and triangle. This relevant information for each
time and space interval is then processed  using a user-defined
operation: a function $f$. The function takes the
filtered result as well as  the  results of previous
iteration steps and combines them. The folding operation provides a way to efficiently analyse larger
data structures in the spatio-temporal continuum.

The signature template of a functional programming-style folding time function is provided below:
{\small
\begin{verbatim}
foldTime[A,T] (
   inv : Invariant, a: A,starttime : T,
   stoptime : T,step: T, 
   f : (A -> Invariant -> A))
\end{verbatim}
}
The function takes our spatio-temporal
model, an invariant {\tt inv} as its first argument. It takes an
initial aggregation value {\tt a} and calls
itself recursively thereby iterating over a time window specified by
{\tt starttime} and {\tt stoptime} using an iteration increase {\tt
  step}. For example, the starttime could be 10 seconds, the stoptime
50 seconds, the step could be 5 seconds. In that case we would filter
the values corresponding to 10, 15, 20, ... , 50 seconds out of the model.
The function selects all relevant information for the iterated time
points and executes a function {\tt f} on it. {\tt f} depends on the
use-case, for example, if we want to count the total solar power generation
in an area, it could sum the values for different PV panels up. The function {\tt f} does take
an aggregated state value {\tt A}. In our example, this could be the
aggregated PV generation for previous time points. In summary, folding can be used to
iterate over spatio-temporal information thereby filtering and
aggregating information by using appropriate instantiations of {\tt f}.

Similarily, the signature of the corresponding fold function for space is provided below:

{\small
\begin{verbatim}
foldSpace[A] (
  inv Invariant,z: A, startarea: Invariant, 
  stoparea: Invariant, steparea: Invariant, 
  f :(A -> Invariant -> A)) 
\end{verbatim}
}
The analysis and aggregation of data is done using a  path in space. From a {\tt
  startarea} to a {\tt stoparea} using a spatial increment {\tt
  steparea}. Analog to time, we iteratively aggregate relevant
information over space into {\tt A} by applying a function {\tt f}.

A main advantage of the folding functions is their genericity, since we can
arbitrarily choose {\tt f}. The functional
approach is  able to handle data in an efficient way: Compilers
can automatically eliminate recursive calls, simplify the
case class-based
data-structures, and parallelization possibilities can  be exploited much
more easily than in standard imperative programming languages.

\subsection{Folding Smart-Grid  Data}

We have instantiated the fold
operations for our spatio-temporal smart grid data. Here, we are
looking at an example where rain-cloud coverage data is aggregated for different
areas. We  analyze moving rain-clouds based on live-weather radar
data and can alert grid operators and
support the assessment of possible implications on solar energy
generation and grid load.

To give a look and feel, Figure~\ref{fig:usercode} shows a simple instantiation of {\tt f}. The aggregated value is
the sum of all rain-cloud covered squares in a geometrical grid.
The example proves that our general functionality can be instantiated
to create customized rule handling for specific applications.

\begin{figure*}
{\small
\begin{verbatim}
def addCloudyArea(total: Int, invariant: Invariant): Int = 
{
  def isCloudyArea(owner: Owner[Any]): Boolean = owner == cloud

  def calculateArea(list: List[Invariant]): Int = {
    val areas: List[Int] = list map {
      inv: Invariant => inv match {
          case IMPLIES(owner: Owner[Any], point: OccupyPoint) => 
             if (isCloudyArea(owner)) 1 else 0
          case IMPLIES(owner: Owner[Any], AND(p1: OccupyPoint, p2: OccupyPoint)) => 
             if (isCloudyArea(owner)) 2 else 0
          case IMPLIES(owner: Owner[Any], BIGAND(points: List[OccupyPoint])) => 
             if (isCloudyArea(owner)) points.length else 0
          case _ => 0
        }
    }
    areas.sum
  }

  val area = invariant match {
    case AND(t1, t2) => calculateArea(t1 :: t2 :: Nil)
    case BIGAND(sublist: List[Invariant]) => calculateArea(sublist)
    case _ => 0
  }

  total + area
}
\end{verbatim}
}
\caption{User-defined code used with BeSpaceD to support spatial
  decisions}
\label{fig:usercode}
\end{figure*}

\section{Smart Energy Relevant Data}
\label{sec:data}
We provide  an overview on relevant data sources for smart
energy systems in this section.
With the advent of advanced mapping techniques, Geographical
Information Systems (GIS), and progress in
visualization, new data sets and means for the classification of data
relevant for smart energy systems have been created. Most importantly
open data  is frequently made available by governments or authentic
agencies and
major parts of relevant data is accessible as open datasets across many
countries. In many cases the 
information is of a spatio-temporal nature. This means information is associated to a location, i.e., identified by
spatial coordinates, for example in terms of latitude and
longitude as well as time. We have identified a listing of relevant data sources
to be of particular relevance for our cause. It can be found in
Table~\ref{tab:list}.
\begin{table*}
{\small
\centering
\begin{tabular}{l l}
\hline
 \textsf{Relevant datasets for Australia} \\
\hline
Australian Bureau of Meteorology & \url{http://www.bom.gov.au/} \\
AREMI &  {\footnotesize \url{http://www.ga.gov.au/scientific-topics/energy/resources/other-renewable-energy-resources/solar-energy}}\\
 &\url{https://data.gov.au/dataset/smart-grid-smart-city-customer-trial-data}\\
\hline
\textsf{Relevant datasets for for India} \\
\hline
Bhuvan (website)&\url{http://bhuvan.nrsc.gov.in/}\\
&\url{https://data.gov.in/catalog/estimated-renewable-energy-potential#web_catalog_tabs_block_10}\\
&\url{https://en.openei.org/datasets/dataset?sectors=smartgrid} \\
\hline
\textsf{Relevant datasets for for South Africa} \\
\hline
Open Energy Database &\url{http://opened.netgen.co.za/} \\
\hline
\textsf{Relevant datasets for for the US} \\
\hline
&\url{http://www.nrdc.org/energy/renewables/energymap.asp} \\
&\url{http://www.nrel.gov/gis/maps.html}\\
&\url{https://www.data.gov/energy/} \\
\hline 
\textsf{Additional Open Data/Source Initiatives} \\
\hline
OpenEnergyMonitor &\url{https://openenergymonitor.org/emon/}\\
Open Energy Data & \url{http://energy.gov/data/open-energy-data} \\
\hline
\end{tabular}
}
\caption{Relevant Datasources}
\label{tab:list}
\end{table*}
 In some cases the information is superimposed on
the country’s map. This is
observed in case of data from India which is  connected to
Bhuvan (Indian Geo-Platform, see table), and Australian data, connected to AREMI: the Australian
Renewable Energy   Mapping Infrastructure (see table). Here, mapping
is done over a national map. Other sources of  datasets include Open
Energy Database from South Africa, while Energy.gov and Data.gov expose open
data from the US. 
Furthermore, there are 
open initiatives and open source approaches that we have investigated 
methods for calculation or estimating the benefits of renewable power
are made available in addition to the data sets by the National Renewable Energy Laboratory 
(NREL) \footnote{\url{http://www.nrel.gov/analysis/data_resources.html}}.
The datasets in data.gov.in include assessment of existing electricity
generation. This is updated year by year. 
Data.gov.in also provides an API. In addition  sources in the XML,
XLS, CSV, JSON and JSONP formats are supported 
for the datasets.  The datasets in AREMI include information about
existing sources of energy such as 
coal based, geothermal, petroleum based, nuclear, as well as renewable sources (e.g.,
hydro, solar, ocean energy, wind, and bio energy sources).  
Furthermore, there are initiatives like 
OpenEI 
that also support APIs such as the
CANData API for exposing information (e.g., fuel economy rate, cost of 
renewable energies, and irradiance). 
For our demonstrator we also rely on live weather data from the
Australian Bureau of Meteorology which is pulled by our framework
using the HTML protocol. 

\subsection*{Possible Usages for Long-term Decision Support}
For the longer term decision support, the presented datasets are
typically used for techno-commercial analysis
to investigate how weak links in grids can be 
strengthened in a location or area. We identify relevant locations and
look at the existing generation in their proximity. As a second step, we  
connect to the irradiance information and weather conditions and infer the possibility of a 
 certain mix of renewable sources to be integrated in that location. Based on the 
strength of renewable energy potential at the location, for example irradiance for solar, speed of wind for 
wind, the possible annual generation can be calculated as a third
step. By using this information, cost and 
incentives for renewable in the location, the Return On Investment (ROI) and the Pay Back Period 
(PBP) for a new concept can be easily deduced as the final result of
our analysis. Overall we derive the
benefits of applying new energy concepts and get an understanding
whether a proposed concept will satisfy the energy needs of the weak
links in the grid. In addition, an understanding of local weather conditions 
provides input on the variability (time of day, time of year,
probability distributions) of the renewable energy
sources. We learn about 
potential implications regarding the amount of reliable power that one
needs to provide and get an
understanding on petroleum and non-renewable basic energy needs for a
location.  This
leads to a suggestion on a possible mix of different energy sources.

\section{The SmartSpace Demonstrator}
\label{sec:demo}

We have built a demonstrator for our short-term decision support
scenario. It is based on the ingredients described in the
previous sections. Our demonstrator realizes the architecture shown in
Figure~\ref{fig:overview}. The BeSpaceD-based decision support is triggered by regularly
recurring feeds of live weather data from the
Australian Bureau of Meteorology. The live weather data is converted
into BeSpaceD models.
Furthermore, the BeSpaceD language is used to describe
rules and models for our smart-grid system. Rules can be changed,
added and removed. This is done by using the Scala/Java-based
development and runtime environment. 
We have experimented with a variety of rules. One is provided below
and realized
using assumption:

\begin{center}
{\small
$t_1 \le time \le t_2$ $\wedge$ \\
 {\textsf{cloud coverage filtered by area}}$_1$ $\ge$ {\textsf{threshold}} $\wedge$
 ... $\wedge$  \\ {\textsf{cloud coverage filtered by area}}$_2$ $\ge$ {\textsf
   threshold} \\
$\longrightarrow$ \\
{\textsf{critical solar energy level}}
}
\end{center}

The rule is given in a very abstract specification level and has the
following form: a condition implies (triggers) a
reaction. In the case above,
 $t_1$ and $t_2$ are timepoints. Thus, the first line specifies a time
 interval. The {\textsf{area}}$_1$ .. {\textsf{area}}$_2$ construct specifies spatial
areas. For example regions on a map identified by their x-- and
y--coordinates. The cloud coverage is calculated for these areas and if
it is above or equal a threshold a {\textsf{critical solar energy
  level}} is assumed to be present. The critical solar energy level
triggers a reaction.
Thus, based on rain-cloud coverage in some areas
stakeholders can be informed through the reaction. The implementation of the rules is written in a less abstract
style. In our case, the specification how the filtering of an area is
done by using the
constructs introduced in the previous sections in BeSpaceD. The above
rule is relatively simple. A system
can comprise a multitude of rules. In addition, one also needs to specify the
triggered reactions, i.e.., XML code. This XML code is then interpreted and
triggers a visualization. 
In addition to the Australian Bureau of Meteorology data, other data sources can be used  in our rules as described
in Section~\ref{sec:data}.

\begin{figure*}
\centering
\includegraphics[width=1\textwidth]{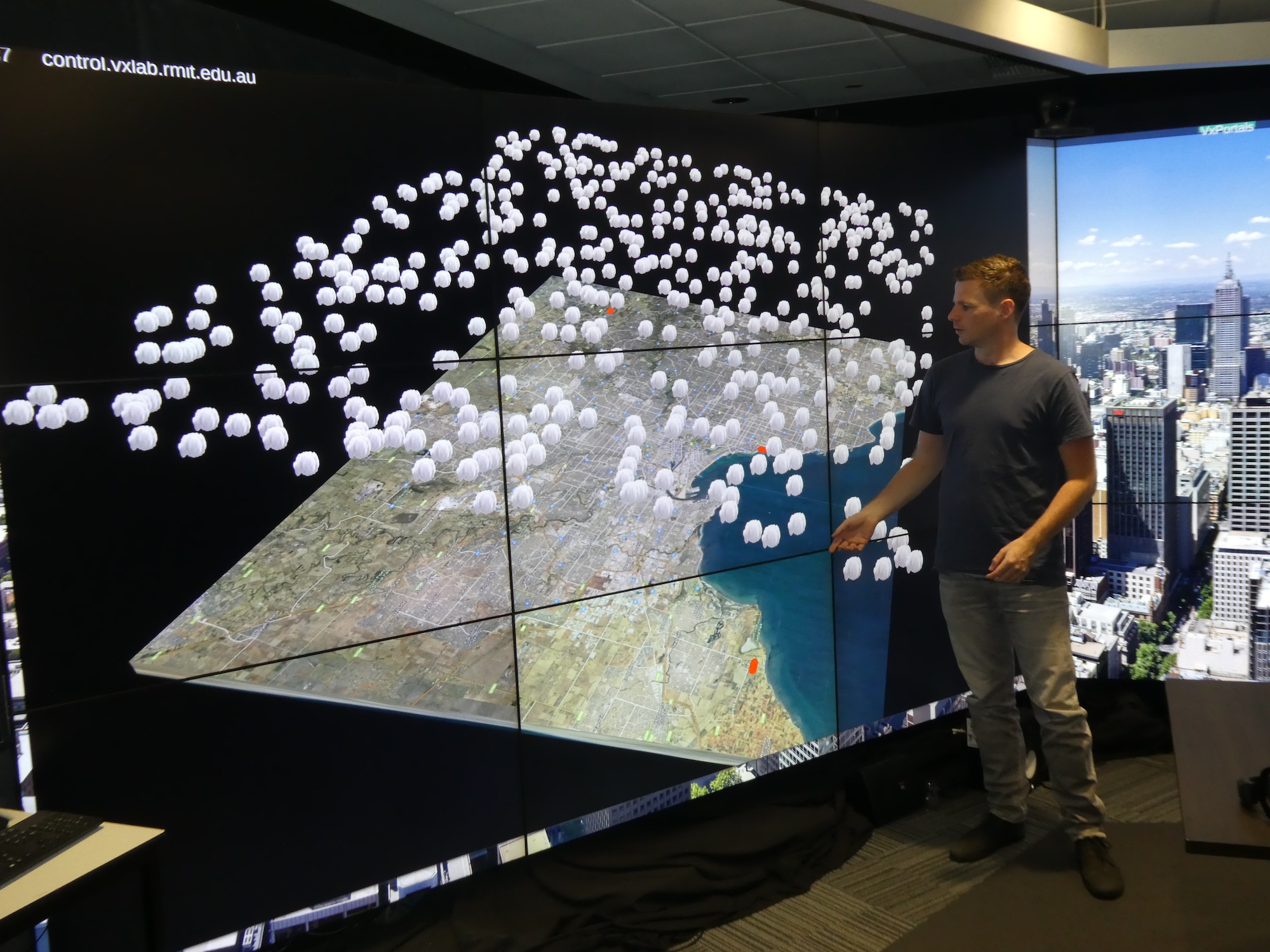} \\ 
~~~~ \\
\includegraphics[width=1\textwidth]{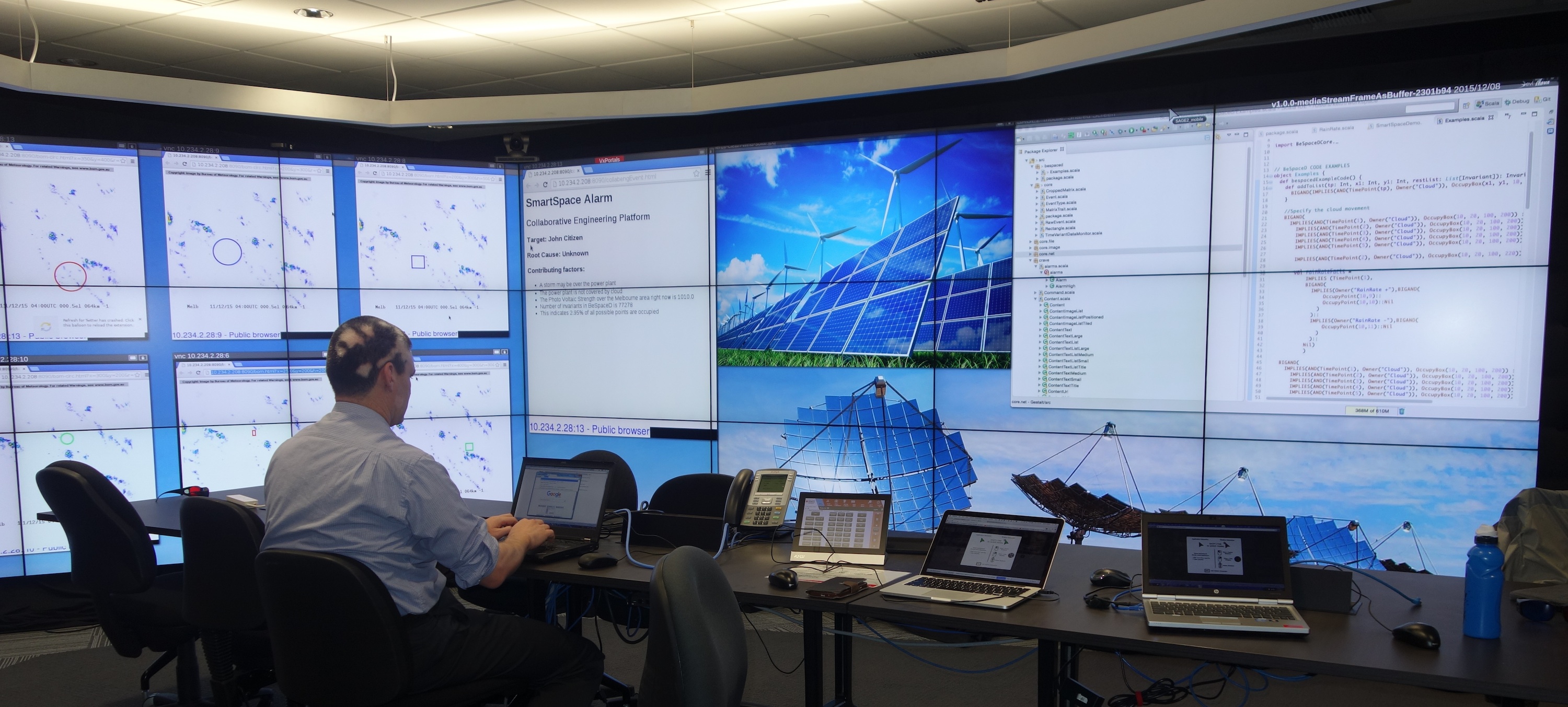}
\caption{Displayed Information in VxLab}
\label{fig:screenshot}
\end{figure*}

Figure~\ref{fig:screenshot} provides two examples of triggered
reaction using the SmartSpace Framework. 
The first picture shows the use of the SmartSpace3D
\cite{smartspace3d} frontend. This view can be triggered through the described
framework. The 3D visualization of relevant objects is based on the BeSpaceD language. In the
picture,  information on rain-cloud coverage, their position in space and
locations of power plants (red dots) are contained. The generated 3D view can be
animated by adding time information to the rain-cloud positions in
BeSpaceD and a head-tracking device can be used to change the views
automatically based on the position of humans in front of the screen.

The second picture shows multiple windows with information triggered
through the SmartSpace framework, such as rain-coverage
information plus augmented data such as highlighting regions of
interest indicated by geometric shapes.
Both, the SmartSpace and SmartSpace3D demonstrators feature a large visualization wall which is made up
of several coupled monitors \cite{vxlab14}. The windows can be resized and moved freely
across the wall \cite{vxlab}. The decision support can also comprise information on
where to put a window on the wall, another feature for 
the user interface. The XML code containing the visualization
information and the infrastructure from
this builds on our collaborative engineering framework and has been
described in \cite{etfa2}. Another form factor that connects to our framework is described in \cite{mobile}. Here, an augmented reality solution for mobile devices connecting to our framework is presented.

\section{Conclusion}
\label{sec:concl}

In this paper, we have presented SmartSpace, a prototype platform for decision
support in the smart energy systems. We have explained our
decision support component BeSpaceD and its operators which have been
implemented to support SmartSpace.  BeSpaceD is able to provide means
to customize decision support rules and allows for an abstraction of
events. Our SmartSpace
framework is also able to allow changes to the rules on-the fly by using
the underlying Scala/Java infrastructure. In addition, we have
motivated and highlighted
the context of our work. We  discussed sources of publicly available
data and have realized and discussed a demonstrator  in this
paper.

Different directions for future work are of interest to
us.: (i) By concentrating on the area of domain specific languages one
could get  an even
more comprehensive
front-end for describing our decision support rules and the underlying
models. This could improve usability and one could imagine to
investigate the human factors of automated decision support.
(ii) The integration of  more data sources into our demonstrator is another
topic we are working on. 
(iii) The study of agent-based approaches for reacting to incidents is
another area of interest. Despite that, in our current work we are primarily concerned with
assisting human stakeholders, not replacing them. (iv) The advancement of our BeSpaceD framework is another
recurring goal.

\subsection*{Acknowledgement}

The authors would like to thank Ian Peake from VxLab and Ed Watkins for their
support during the implementation of the SmartSpace project.

\bibliographystyle{plain}

\end{document}